# Graphical-model Based Multiple Testing under Dependence, with Applications to Genome-wide Association Studies

**Jie Liu**
Computer Sciences, UW-Madison

**Chunming Zhang**
Statistics, UW-Madison

**Catherine McCarty**
Essentia Institute of Rural Health

**Peggy Peissig**
Marshfield Clinic Research Foundation

**Elizabeth Burnside**
Radiology, UW-Madison

**David Page**
BMI & CS, UW-Madison

## Abstract

Large-scale multiple testing tasks often exhibit dependence, and leveraging the dependence between individual tests is still one challenging and important problem in statistics. With recent advances in graphical models, it is feasible to use them to perform multiple testing under dependence. We propose a multiple testing procedure which is based on a Markov-random-field-coupled mixture model. The ground truth of hypotheses is represented by a latent binary Markov random field, and the observed test statistics appear as the coupled mixture variables. The parameters in our model can be automatically learned by a novel EM algorithm. We use an MCMC algorithm to infer the posterior probability that each hypothesis is null (termed *local index of significance*), and the false discovery rate can be controlled accordingly. Simulations show that the numerical performance of multiple testing can be improved substantially by using our procedure. We apply the procedure to a real-world genome-wide association study on breast cancer, and we identify several SNPs with strong association evidence.

## 1 Introduction

Observations from large-scale multiple testing problems often exhibit dependence. For instance, in genome-wide association studies, researchers collect hundreds of thousands of highly correlated genetic markers (single-nucleotide polymorphisms, or SNPs) with the purpose of identifying the subset of markers associated with a heritable disease or trait. In functional magnetic resonance imaging studies of the brain, thousands of spatially correlated voxels are collected while subjects are performing certain tasks, with the purpose of detecting the relevant voxels. The most popular family of large-scale multiple testing procedures is the false discovery rate analysis, such as the $p$-value thresholding procedures (Benjamini & Hochberg, 1995, 2000; Genovese & Wasserman, 2004), the local false discovery rate procedure (Efron et al., 2001), and the positive false discovery rate procedure (Storey, 2002, 2003). However, all these classical multiple testing procedures ignore the correlation structure among the individual factors, and the question is *whether we can reduce the false non-discovery rate by leveraging the dependence, while still controlling the false discovery rate in multiple testing.*

Graphical models provide an elegant way of representing dependence. With recent advances in graphical models, especially more efficient algorithms for inference and parameter learning, it is feasible to use these models to leverage the dependence between individual tests in multiple testing problems. One influential paper (Sun & Cai, 2009) in the statistics community uses a hidden Markov model to represent the dependence structure, and has shown its optimality under certain conditions and its strong empirical performance. It is the first graphical model (and the only one so far) used in multiple testing problems. However, their procedure can only deal with a sequential dependence structure, and the dependence parameters are homogenous. *In this paper, we propose a multiple testing procedure based on a Markov-random-field-coupled mixture model which allows arbitrary dependence structures and heterogeneous dependence parameters*. This extension requires more sophisticated algorithms for parameter learning and inference. For parameter learning, we design an EM algorithm with MCMC in the E-step and persistent contrastive divergence algorithm (Tieleman, 2008) in the M-step. We use the MCMC algorithm to infer the posterior probability that each hypothesis is null (termed *local index of significance* or LIS). Finally, the false discovery rate can be controlled by thresholding the LIS.

Section 2 introduces related work and our procedure. Sections 3 and 4 evaluate our procedure on a variety of simulations, and the empirical results show that the numerical performance can be improved substantially by using our procedure. In Section 5, we apply the procedure to a real-world genome-wide association study (GWAS) on breast cancer, and we identify several SNPs with strong association evidence. We finally conclude in Section 6.

## 2 Method

### 2.1 Terminology and Previous Work

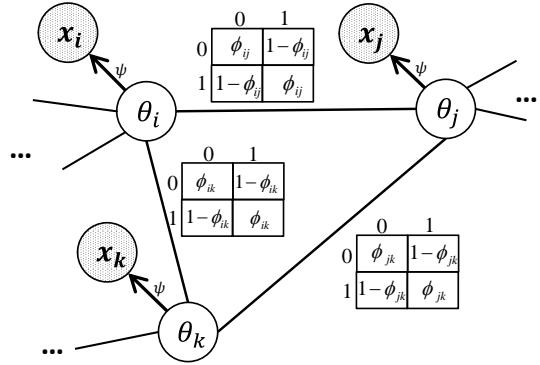

Figure 1: The MRF-coupled mixture model for three dependent hypotheses $\mathcal{H}_i$, $\mathcal{H}_j$ and $\mathcal{H}_k$ with *observed* test statistics ($x_i$, $x_j$ and $x_k$) and *latent* ground truth ($\theta_i, \theta_j$ and $\theta_k$). The dependence is captured by potential functions parameterized by $\phi_{ij}, \phi_{jk}$ and $\phi_{ik}$, and coupled mixtures are parameterized by $\psi$.

Table 1: Classification of tested hypotheses

|  | Not rejected | Rejected | Total |
|---|---|---|---|
| Null | $N_{00}$ | $N_{10}$ | $m_0$ |
| Non-null | $N_{01}$ | $N_{11}$ | $m_1$ |
| Total | $S$ | $R$ | $m$ |

Suppose that we carry out $m$ tests whose results can be categorized as in Table 1. *False discovery rate* (FDR), defined as $E(N_{10}/R|R > 0)P(R > 0)$, depicts the expected proportion of incorrectly rejected null hypotheses (Benjamini & Hochberg, 1995). *False non-discovery rate* (FNR), defined as $E(N_{01}/S|S > 0)P(S > 0)$, depicts the expected proportion of false non-rejections in those tests whose null hypotheses are not rejected (Genovese & Wasserman, 2002). An FDR procedure is *valid* if it controls FDR at a nominal level, and *optimal* if it has the smallest FNR among all the valid FDR procedures (Sun & Cai, 2009). The effects of correlation on multiple testing have been discussed, under different assumptions, with a focus on the validity issue (Benjamini & Yekutieli, 2001; Finner & Roters, 2002; Owen, 2005; Sarkar, 2006; Efron, 2007; Farcomeni, 2007; Romano et al., 2008; Wu, 2008; Blanchard & Roquain, 2009). The efficiency issue has also been investigated (Yekutieli & Benjamini, 1999; Genovese et al., 2006; Benjamini & Heller, 2007; Zhang et al., 2011), indicating FNR could be decreased by considering dependence in multiple testing. Several approaches have been proposed, such as dependence kernels (Leek & Storey, 2008), factor models (Friguet et al., 2009) and principal factor approximation (Fan et al., 2012). Sun & Cai (2009) explicitly use a hidden Markov model (HMM) to represent the dependence structure and analyze the optimality under the compound decision framework (Sun & Cai, 2007). However, their procedure can only deal with sequential dependence, and it uses only a single dependence parameter throughout. In this paper, we replace HMM with a Markov-random-field-coupled mixture model, which allows richer and more flexible dependence structures. The Markov-random-field-coupled mixture models are related to the hidden Markov random field models used in many image segmentation problems (Zhang et al., 2001; Celeux et al., 2003; Chatzis & Varvarigou, 2008).

### 2.2 Our Multiple Testing Procedure

Let $\mathbf{x} = (x_1, ..., x_m)$ be a vector of test statistics from a set of hypotheses $(\mathcal{H}_1, ..., \mathcal{H}_m)$. The ground truth of these hypotheses is denoted by a latent Bernoulli vector $\boldsymbol{\theta} = (\theta_1, ..., \theta_m) \in \{0,1\}^m$, with $\theta_i = 0$ denoting that the hypothesis $\mathcal{H}_i$ is null and $\theta_i = 1$ denoting that the hypothesis $\mathcal{H}_i$ is non-null. The dependence among these hypotheses is represented as a binary Markov random field (MRF) on $\boldsymbol{\theta}$. The structure of the MRF can be described by an undirected graph $\mathcal{G}(\mathcal{V}, \mathcal{E})$ with the node set $\mathcal{V}$ and the edge set $\mathcal{E}$. The dependence between $\mathcal{H}_i$ and $\mathcal{H}_j$ is denoted by an edge connecting node$_i$ and node$_j$ in $\mathcal{E}$, and the strength of dependence is parameterized by the potential function on the edge. Suppose that the probability density function of the test statistic $x_i$ given $\theta_i = 0$ is $f_0$, and the density of $x_i$ given $\theta_i = 1$ is $f_1$. Then, $\mathbf{x}$ is an MRF-coupled mixture. The mixture model is parameterized by a parameter set $\vartheta = (\boldsymbol{\phi}, \boldsymbol{\psi})$, where $\boldsymbol{\phi}$ parameterizes the binary MRF and $\boldsymbol{\psi}$ parameterizes $f_0$ and $f_1$. For example, if $f_0$ is standard normal $\mathcal{N}(0,1)$ and $f_1$ is noncentered normal $\mathcal{N}(\mu, 1)$, then $\boldsymbol{\psi}$ only contains parameter $\mu$. Figure 1 shows the MRF-coupled mixture model for three dependent hypotheses $\mathcal{H}_i$, $\mathcal{H}_j$ and $\mathcal{H}_k$.

In our MRF-coupled mixture model, $\mathbf{x}$ is observable, and $\boldsymbol{\theta}$ is hidden. With the parameter set $\vartheta = (\boldsymbol{\phi}, \boldsymbol{\psi})$, the joint probability density over $\mathbf{x}$ and $\boldsymbol{\theta}$ is

$$P(\mathbf{x}, \boldsymbol{\theta}|\boldsymbol{\phi}, \boldsymbol{\psi}) = P(\boldsymbol{\theta}; \boldsymbol{\phi}) \prod_{i=1}^{m} P(x_i|\theta_i; \boldsymbol{\psi}). \quad (1)$$

Define the marginal probability that $\mathcal{H}_i$ is null given all the observed statistics $\mathbf{x}$ under the parameters in $\vartheta$, $P_\vartheta(\theta_i = 0|\mathbf{x})$, to be the *local index of significance* (LIS) for $\mathcal{H}_i$ (Sun & Cai, 2009). If we can accurately calculate the posterior marginal probabilities of $\boldsymbol{\theta}$ (or LIS), then we can use a step-up procedure to control FDR at the nominal level $\alpha$ as follows (Sun & Cai, 2009). We first sort LIS from the smallest value to the largest value. Suppose $\text{LIS}_{(1)}$, $\text{LIS}_{(2)}$, ..., and $\text{LIS}_{(m)}$ are the ordered LIS, and the corresponding hypotheses are $\mathcal{H}_{(1)}$, $\mathcal{H}_{(2)}$,..., and $\mathcal{H}_{(m)}$. Let

$$k = \max\left\{i : \frac{1}{i}\sum_{j=1}^{i} \text{LIS}_{(j)} \leq \alpha\right\}. \qquad (2)$$

Then we reject $\mathcal{H}_{(i)}$ for $i = 1, ..., k$.

Therefore, the key inferential problem that we need to solve is that of computing the posterior marginal distribution of the hidden variables $\theta_i$ given the test statistics $\mathbf{x}$, namely $P_\vartheta(\theta_i = 0|\mathbf{x})$, for $i = 1, ..., m$. It is a typical inference problem if the parameters in $\vartheta$ are known. Section 2.3 provides possible inference algorithms for calculating $P_\vartheta(\theta_i = 0|\mathbf{x})$ for given $\vartheta$. However, $\vartheta$ is usually unknown in real-world applications, and we need to estimate it. Section 2.4 provides a novel EM algorithm for parameter learning in our MRF-coupled mixture model.

### 2.3 Posterior Inference

Now we are interested in calculating $P_\vartheta(\theta_i = 0|\mathbf{x})$ for a given parameter set $\vartheta$. One popular family of inference algorithms is the sum-product family (Kschischang et al., 2001), also known as belief propagation (Yedidia et al., 2000). For loop-free graphs, belief propagation algorithms provide exact inference results with a computational cost linear in the number of variables. In our MRF-coupled mixture model, the structure of the latent MRF is described by a graph $\mathcal{G}(\mathcal{V}, \mathcal{E})$. When $\mathcal{G}$ is chain structured, the instantiation of belief propagation is the forward-backward algorithm (Baum et al., 1970). When $\mathcal{G}$ is tree structured, the instantiation of belief propagation is the upward-downward algorithm (Crouse et al., 1998). For graphical models with cycles, loopy belief propagation (Murphy et al., 1999; Weiss, 2000) and the tree-reweighted algorithm (Wainwright et al., 2003a) can be used for approximate inference. Other inference algorithms for graphical models include junction trees (Lauritzen & Spiegelhalter, 1988), sampling methods (Gelfand & Smith, 1990), and variational methods (Jordan et al., 1999). Recent papers (Schraudolph & Kamenetsky, 2009; Schraudolph, 2010) discuss exact inference algorithms on binary Markov random fields which allow loops. In our simulations, we use belief propagation when the graph $\mathcal{G}$ has no loops. When $\mathcal{G}$ has loops (e.g. in the simulations on genetic data and the real-world application), we use a Markov chain Monte Carlo (MCMC) algorithm to perform inference for $P_\vartheta(\theta_i = 0|\mathbf{x})$.

### 2.4 Parameters and Parameter Learning

In our procedure, the dependence among these hypotheses is represented by a graphical model on the latent vector $\boldsymbol{\theta}$ parameterized by $\boldsymbol{\phi}$, and observed test statistics $\mathbf{x}$ are represented by the coupled mixture parameterized by $\boldsymbol{\psi}$. In Sun and Cai's work on HMMs, $\boldsymbol{\phi}$ is the transition parameter and $\boldsymbol{\psi}$ is the emission parameter. One implicit assumption in their work is that the transition parameter and the emission parameter stay the same for $i(i = 1, ..., m)$. Our extension to MRFs also allows us to untie these parameters. In the second set of basic simulations in Section 3, we make $\boldsymbol{\phi}$ and $\boldsymbol{\psi}$ heterogeneous and investigate how this affects the numerical performance. In the simulations on genetic data in Section 4 and the real-world GWAS application in Section 5, we have different parameters for SNP pairs with different levels of correlation.

In our model, learning $(\boldsymbol{\phi}, \boldsymbol{\psi})$ is difficult for two reasons. First, learning parameters is difficult by nature in undirected graphical models due to the global normalization constant (Wainwright et al., 2003b; Welling & Sutton, 2005). State-of-the-art MRF parameter learning methods include MCMC-MLE (Geyer, 1991), contrastive divergence (Hinton, 2002) and variational methods (Ganapathi et al., 2008). Several new sampling methods with higher efficiency have been recently proposed, such as persistent contrastive divergence (Tieleman, 2008), fast-weight contrastive divergence (Tieleman & Hinton, 2009), tempered transitions (Salakhutdinov, 2009), and particle-filtered MCMC-MLE (Asuncion et al., 2010). In our procedure, we use the persistent contrastive divergence algorithm to estimate parameters $\boldsymbol{\phi}$. Another difficulty is that $\boldsymbol{\theta}$ is latent and we only have one observed training sample $\mathbf{x}$. We use an EM algorithm to solve this problem. In the E-step, we run our MCMC algorithm in Section 2.3 to infer the latent $\boldsymbol{\theta}$ based on the currently estimated parameters $\vartheta = (\boldsymbol{\phi}, \boldsymbol{\psi})$. In the M-step, we run the persistent contrastive divergence (PCD) algorithm (Tieleman, 2008) to estimate $\boldsymbol{\phi}$ from the currently inferred $\boldsymbol{\theta}$. Note that PCD is also an iterative algorithm, and we run it until it converges in each M-step. In the M-step, we also do a maximum likelihood estimation of $\boldsymbol{\psi}$ from the currently inferred $\boldsymbol{\theta}$ and observed $\mathbf{x}$. We run the EM algorithm until both $\boldsymbol{\phi}$ and $\boldsymbol{\psi}$ converge. Although this EM algorithm involves intensive computation in both E-step and M-step, it converges very quickly in our experiments.

# 3 Basic Simulations

In the basic simulations, we investigate the numerical performance of our multiple testing approach on different fabricated dependence structures where we can control the ground truth parameters. We first simulate $\boldsymbol{\theta}$ from $P(\boldsymbol{\theta}; \boldsymbol{\phi})$ and then simulate $\mathbf{x}$ from $P(\mathbf{x}|\boldsymbol{\theta}; \boldsymbol{\psi})$ under a variety of settings of $\vartheta = (\boldsymbol{\phi}, \boldsymbol{\psi})$. Because we have the ground truth parameters, we have two versions of our multiple testing approach, namely the oracle procedure (OR) and the data-driven procedure (LIS). The oracle procedure knows the true parameters $\vartheta$ in the graphical models, whereas the data-driven procedure does not and has to estimate $\vartheta$. The baseline procedures include the BH procedure (Benjamini & Hochberg, 1995) and the adaptive $p$-value procedure (AP) (Benjamini & Hochberg, 2000; Genovese & Wasserman, 2004) which are compared by Sun & Cai (2009). We include another baseline procedure, the local false discovery rate procedure (localFDR) (Efron et al., 2001). The adaptive $p$-value procedure requires a consistent estimate of the proportion of the true null hypotheses. The localFDR procedure requires a consistent estimate of the proportion of the true null hypotheses and the knowledge of the distribution of the test statistics under the null and under the alternative. In our simulations, we endow AP and localFDR with the ground truth values of these in order to let these baseline procedures achieve their best performance.

In the simulations, we assume that the observed $x_i$ under the null hypothesis (namely $\theta_i = 0$) is standard-normally distributed and that $x_i$ under the alternative hypothesis (namely $\theta_i = 1$) is normally distributed with mean $\mu$ and standard deviation 1.0. We choose the setup and parameters to be consistent with the work of Sun & Cai (2009) when possible. In total, we consider three MRF models, namely a chain-structured MRF, tree-structured MRF and grid-structured MRF. For chain-MRF, we choose the number of hypotheses $m = 3,000$. For tree-MRF, we choose perfect binary trees of height 12 which yields a total number of $8,191$ hypotheses. For grid-MRF, we choose the number of rows and the number of columns to be 100 which yields a total number of $10,000$ hypotheses. In all the experiments, we choose the number of replications $N = 500$ which is also the same as the work of Sun & Cai (2009). In total, we have three sets of simulations with different goals as follows.

**Basic simulation 1:** We stay consistent with Sun & Cai (2009) in the simulations except that we use the three MRF models. In all three structures, $(\theta_i)_1^m$ is generated from the MRFs whose potentials on the edges are $\begin{pmatrix} \phi & 1-\phi \\ 1-\phi & \phi \end{pmatrix}$. Therefore, $\boldsymbol{\phi}$ only contains parameter $\phi$, and $\boldsymbol{\psi}$ only includes parameter $\mu$.

**Basic simulation 2:** One assumption in basic simulation 1 is that the parameters $\phi$ and $\mu$ are homogeneous in the sense that they stay the same for $i(i = 1, ..., m)$. This assumption is carried down from the work of Sun & Cai (2009). However in many real-world applications, the transition parameters can be different across the multiple hypotheses. Similarly, the test statistics for the non-null hypotheses, although normally distributed and standardized, could have different $\mu$ values. Therefore, we investigate the situation where the parameters can vary in different hypotheses. The simulations are carried out for all three different dependence structures aforementioned. In the first set of simulations, instead of fixing $\phi$, we choose $\phi$'s uniformly distributed on the interval $(0.8 - \Delta(\phi)/2, 0.8 + \Delta(\phi)/2)$. In the second set of simulations, instead of fixing $\mu$, we choose $\mu$'s uniformly distributed on the interval $(2.0 - \Delta(\mu)/2, 2.0 + \Delta(\mu)/2)$. The oracle procedure knows the true parameters. The data-driven procedure does not know the parameters, and assumes the parameters are homogeneous.

**Basic simulation 3:** Another implicit assumption in basic simulation 1 is that each individual test in the multiple testing problem is exact. Many widely used hypothesis tests, such as Pearson's $\chi^2$ test and the likelihood ratio test, are asymptotic in the sense that we only know the limiting distribution of the test statistics for large samples. As an example, we simulate the two-proportion $z$-test in this section and show how the sample size affects the performance of the procedures when the individual test is asymptotic. Suppose that we have $n$ samples (half of them are positive samples and half of them are negative samples). For each sample, we have $m$ Bernoulli distributed attributes. A fraction of the attributes are relevant. If the attribute $A$ is relevant, then the probability of "heads" in the positive samples ($p_A^+$) is different from that in the negative samples ($p_A^-$). $p_A^+$ and $p_A^-$ are the same if $A$ is non-relevant. For each individual test, the null hypothesis is that the attribute is not relevant, and the alternative hypothesis is otherwise. The two-proportion $z$-test can be used to test whether $p_A^+ - p_A^-$ is zero, which yields an asymptotic $\mathcal{N}(0, 1)$ under the null and $\mathcal{N}(\mu, 1)$ under the alternative ($\mu$ is nonzero). In the simulations, we fix $\mu$, but vary the sample size $n$, and apply the aforementioned tree-MRF structure ($m = 8,191$). The oracle procedure and localFDR only know the limiting distribution of the test statistics and assume the test statistics exactly follow the limiting distributions even when the sample size is small.

Figure 2 shows the numerical results in basic simulation 1. Figures (1a)-(1f) are for the chain structure. Figures (2a)-(2f) are for tree structure. Figures (3a)-(3f) are for the grid structure. In Figures (1a)-(1c), (2a)-(2c) and (3a)-(3c), we set $\mu = 2$ and plot FDR, FNR and the average number of true positives (ATP) when we vary $\phi$ between 0.2 and 0.8. In Figures (1d)-(1f), (2d)-(2f) and (3d)-(3f), we set $\phi = 0.8$ and plot FDR, FNR and ATP when we vary $\mu$ between 1.0 and 4.0. The nominal FDR level is set to be 0.10. From Figure 2, we can observe comparable numerical results between the chain structure and tree structure. The FDR levels of all five procedures are controlled at 0.10 and BH is conservative. From the plots for FNR and ATP, we can observe that the data-driven procedure performs almost the same as the oracle procedure, and they dominate the $p$-value thresholding procedures BH and AP. The oracle procedure and the data-driven procedure also dominate localFDR except when $\phi = 0.5$, when they perform comparably. This is to be expected because the dependence structure is no longer informative when $\phi$ is 0.5. In this situation when the hypotheses are independent, our procedure reduces to the localFDR procedure. As $\phi$ departs from 0.5 and approaches either 0 or 1.0, the difference between OR/LIS and the baselines gets larger. When the individual hypotheses are easy to test (large $\mu$ values), the differences between them are not substantial. When we turn to the grid structure, the numerical performance is similar to that in the chain structure and the tree structure except for two observations. First, the data-driven procedure does not appear to control the FDR at 0.1 when $\mu$ is small (e.g. $\mu = 1.0$), although the oracle procedure does, which indicates the parameter estimation in the EM algorithm is difficult when $\mu$ is small. In other words, with a limited number of hypotheses, it is difficult to estimate the pairwise potential parameters if the test statistics of the non-nulls do not look much different from the test statistics of the nulls. The second observation is that the slopes of the FNR curve and ATP curve for the grid structure are different from those in the chain and tree structures. The reason is that the connectivity in the grid structure is higher than that in the chain and tree. Therefore we can observe that even when the individual hypotheses are difficult to test (small $\mu$ values), the FNR is still low because each individual hypothesis has more neighbors in the grid than in the chain or tree, and the neighbors are informative.

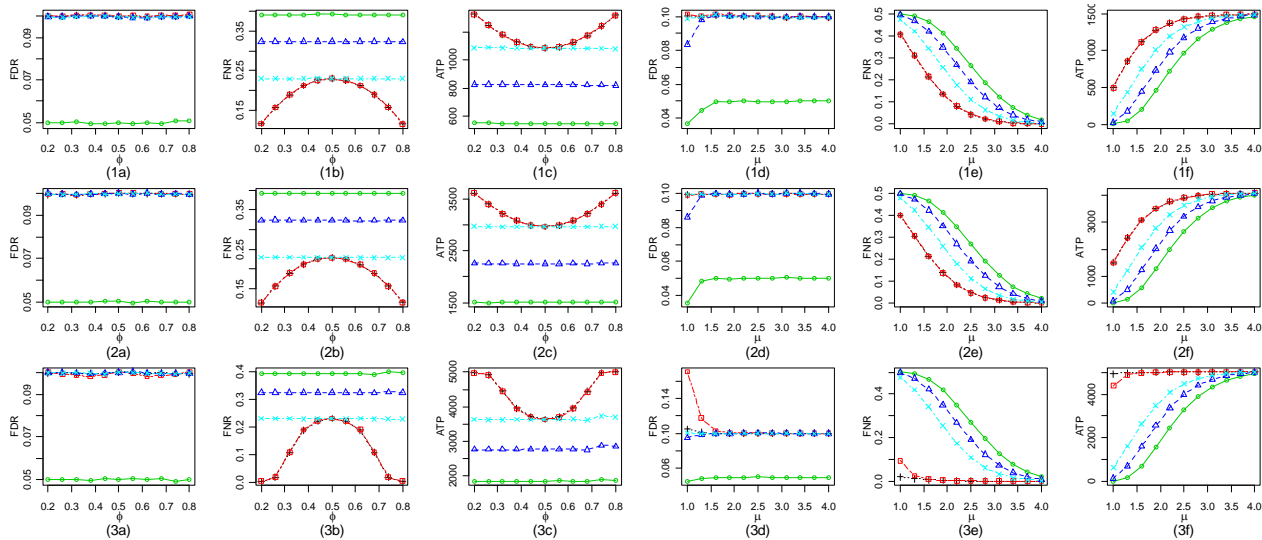

Figure 2: Comparison of BH(○), AP(△), localFDR(×), OR (+), and LIS (□) in basic simulation 1: (1) chain-MRF, (2) tree-MRF, (3) grid-MRF; (a) FDR vs $\phi$, (b) FNR vs $\phi$, (c) ATP vs $\phi$, (d) FDR vs $\mu$, (e) FNR vs $\mu$, (f) ATP vs $\mu$.

Figure 3 shows the numerical performance in basic simulation 2. Figures (1a)-(1f), (2a)-(2f), and (3a)-(3f) correspond to the chain structure, the tree structure and the grid structure respectively. In Figures (1a)-(1c), (2a)-(2c), and (3a-3c), we set $\mu = 2$ and vary $\Delta(\phi)$ between 0 and 0.4. In Figures (1d)-(1f), (2d)-(2f), and (3d)-(3f), we set $\phi = 0.8$ and vary $\Delta(\mu)$ between 0 and 4.0. Again, the nominal FDR level is set to be 0.10. From Figure 3, we observe that all five procedures control FDR at the nominal level and BH is conservative when the transition parameter $\phi$ is heterogeneous. However, the data-driven procedure becomes more and more conservative as we increase the variance of $\phi$ in the grid-structure. Nevertheless, the data-driven procedure does not lose much efficiency compared with the oracle procedure based on FNR and ATP. Both the data-driven procedure and the oracle procedure dominate the three baselines. When the $\mu$ parameter is heterogeneous, all five procedures are still valid, but the data-driven procedure becomes more and more conservative as we increase the variance of $\mu$. The data-driven procedure can be more conservative than the BH procedure when $\Delta(\mu)$ is large enough. The conservativeness appears most severe in the grid-structure. However when we look at the FNR and ATP, the data-driven procedure still dominates BH, AP and localFDR substantially in all the situations, although the data-driven procedure loses a certain amount of efficiency compared with the oracle procedure when the variance of $\mu$ gets large.

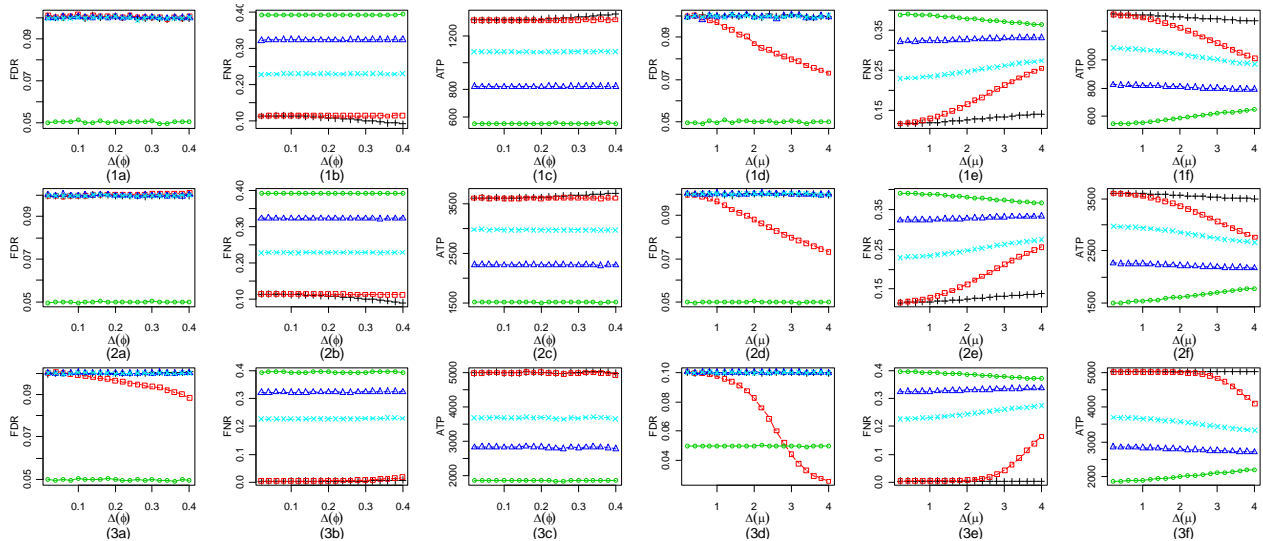

Figure 3: Comparison of BH($\bigcirc$), AP($\triangle$), localFDR($\times$), OR (+), and LIS ($\square$) in basic simulation 2: (1) chain-MRF, (2) tree-MRF, (3) grid-MRF; (a) FDR vs $\Delta(\phi)$, (b) FNR vs $\Delta(\phi)$, (c) ATP vs $\Delta(\phi)$, (d) FDR vs $\Delta(\mu)$, (e) FNR vs $\Delta(\mu)$, (f) ATP vs $\Delta(\mu)$.

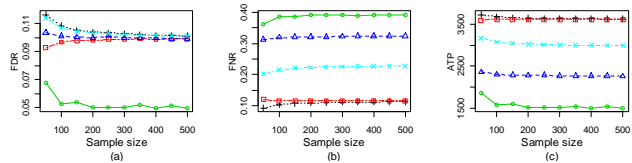

Figure 4: Comparing BH($\bigcirc$), AP($\triangle$), localFDR($\times$), OR(+), and LIS($\square$) in basic simulation 3: (a)FDR vs $n$, (b)FNR vs $n$, (c)ATP vs $n$.

Figure 4 shows the results from basic simulation 3. The oracle procedure and localFDR are liberal when the sample size is small. This is because when the sample size is small, there exists a discrepancy between the true distribution of the test statistic and the limiting distribution. Quite surprisingly, the data-driven procedure stays valid. The reason is that the data-driven procedure can estimate the parameters from data. The data-driven procedure and the oracle procedure still have comparable performance and enjoy a much lower level of FNR compared with the baselines. For all the basic simulations, we set the nominal FDR level to be 0.10. We have also replicated the basic simulations by setting the nominal level to be 0.05, and similar conclusions can be made.

## 4 Simulations on Genetic Data

Unlike the fabricated dependence structures in the basic simulations in Section 3, the dependence structure in the simulations on genetic data in this section is real. We simulate the linkage disequilibrium structure of a segment on human chromosome 22, and treat a test of whether a SNP is associated as one individual test. We follow the simulation settings in the work of Wu et al. (2010). We use HAPGEN2 (Su et al., 2011) and the CEU sample of HapMap (The International HapMap Consortium, 2003) (Release 22) to generate SNP genotype data at each of the $2,420$ loci between bp 14431347 and bp 17999745 on Chromosome 22. A total of 685 out of $2,420$ SNPs can be genotyped with the Affymetrix 6.0 array. These are the typed SNPs that we use for our simulations. Within the overall $2,420$ SNPs, we randomly select 10 SNPs to be the *causal* SNPs. All the SNPs on the Affymetrix 6.0 array whose $r^2$ values, according to HapMap, with any of the causal SNPs are above $t$ are set to be the *associated* SNPs. In the simulations, we report results for three different $t$ values, namely 0.8, 0.5 and 0.25. We also simulate three different genetic models (additive model, dominant model, and recessive model) with different levels of relative risk (1.2 and 1.3). In total, we simulate 250 cases and 250 controls. The experiment is replicated for 100 times and the average result is provided. With the simulated data, we apply our multiple testing procedure (LIS) and three baseline procedures: the BH procedure, the adaptive $p$-value procedure (AP), and the local false discovery rate procedure (localFDR). Because the dependence structure is real and the ground truth parameters are unknown to us, we do not have the oracle procedure in the simulations on genetic data.

With the simulated genetic data, we use two commonly used tests in genetic association studies, namely two-proportion $z$-test and Cochran-Armitage's trend test (CATT) (Cochran, 1954; Armitage, 1955; Slager & Schaid, 2001; Freidlin et al., 2002) as the individual tests for the association of each SNP. CATT also yields an asymptotic $\mathcal{N}(0,1)$ under the null and $\mathcal{N}(\mu,1)$ under the alternative ($\mu$ is nonzero). Therefore, we parameterize $\boldsymbol{\psi} = (\mu_1, \sigma_1^2)$ where $\mu_1$ and $\sigma_1^2$ are the mean and variance of the test statistics under alternative. The graph structure is built as follows. Each SNP becomes a node in the graph. For each SNP, we connect it with the SNP with the highest $r^2$ value with it. There are in total 490 edges in the graph. We further categorize the edges into a high correlation edge set $\mathcal{E}_h$ ($r^2$ above 0.8), medium correlation edge set $\mathcal{E}_m$ ($r^2$ between 0.5 and 0.8) and low correlation edge set $\mathcal{E}_l$ ($r^2$ between 0.25 and 0.5). We have three different parameters ($\phi_h$, $\phi_m$, and $\phi_l$) for the three sets of edges. Then the density of $\boldsymbol{\theta}$ in formula (1) takes the form

$$P(\boldsymbol{\theta}; \boldsymbol{\phi}) \propto \exp\{ \sum_{(i,j) \in \mathcal{E}_h} \phi_h I(\theta_i = \theta_j) + \sum_{(i,j) \in \mathcal{E}_m} \phi_m I(\theta_i = \theta_j) + \sum_{(i,j) \in \mathcal{E}_l} \phi_l I(\theta_i = \theta_j)\}, \quad (3)$$

where $I(\theta_i = \theta_j)$ is an indicator variable that indicates whether $\theta_i$ and $\theta_j$ take the same value. In the MCMC algorithm, we run the Markov chain for 20,000 iterations with a burn-in of 100 iterations. In the PCD algorithm, we generate 100 particles. In each iteration of PCD learning, the particles move forward for 5 iterations (the $n$ parameter in PCD-$n$). The learning rate in PCD gradually decreases as suggested by Tieleman (2008). The EM algorithm converges after about 10 to 20 iterations, which usually take less than 10 minutes on a 3.00GHz CPU.

Figure 5 shows the performance of the procedures in the additive models with the homozygous relative risk set to 1.2 and 1.3. The test statistics are from a two-proportion $z$-test. We have also replicated the simulations on Cochran-Armitage's trend test, and the results are almost the same. In Figure 5, table (1) summarizes the empirical FDR and the total number of true positives (#TP) of our LIS procedure, BH, AP and localFDR (lfdr), in the additive models with different (homozygous) relative risk levels, when we vary $t$ and when we vary the nominal FDR level $\alpha$. We regard a SNP having $r^2$ above $t$ with any causal SNP as an associated SNP, and we regard a rejection of the null hypothesis for an associated SNP as a true positive. Our LIS procedure and localFDR are valid while being conservative. BH and AP appear liberal in some of the configurations. In any of the circumstances, our LIS procedure can identify more associated SNPs than the baselines. We can find a clue to why our procedure LIS is being conservative from the results in Figure 3. In basic simulation 2, we observe that when the parameters $\mu$ and $\phi$ are heterogeneous and we carry out the data-driven procedure under the homogeneous parameter assumption, the data-driven procedure is conservative. The discrepancy between the nominal FDR level and the empirical FDR level increases as the parameters move further away from homogeneity. Although we assign three different parameters $\phi_h$, $\phi_m$, and $\phi_l$ to $\mathcal{E}_h$, $\mathcal{E}_m$ and $\mathcal{E}_l$ respectively, the edges within the same set (e.g. $\mathcal{E}_l$) may still be heterogeneous. The fact that the LIS procedure recaptures more true positives than the baselines while remaining more conservative in many configurations indicates that the local indices of significance provide a ranking more efficient than the ranking provided by the $p$-values from the individual tests. Therefore, we further plot the ROC curves and precision-recall (PR) curves when we rank SNPs by LIS and by the $p$-values from the two-proportion $z$-test. The ROC curve and PR curve are vertically averaged from 100 replications. Subfigures (2a)-(2f) are for the additive model with homozygous relative risk level set to be 1.2. Subfigures (3a)-(3f) are for the additive model with homozygous relative risk level set to be 1.3. It is observed that the curves from LIS dominate those from the $p$-values from individual tests in most places, which further suggests that LIS provides a more efficient ranking of the SNPs than the individual tests.

Figure 6 shows the performance of the procedures in the dominant model and the recessive model with the homozygous relative risk set to be 1.2. The test statistics are from a two-proportion $z$-test. In Figure 6, table (1) summarizes the empirical FDR and the total number of true positives (#TP) of our LIS procedure, BH, AP and localFDR (lfdr) in the dominant model and the recessive model when we vary $t$ and when we vary the nominal FDR level $\alpha$. Our LIS procedure and localFDR are valid while being conservative in all configurations, and they appear more conservative in the recessive model than in the dominant model. On the other hand, BH and AP appear liberal in the recessive model. Our LIS procedure still confers an advantage

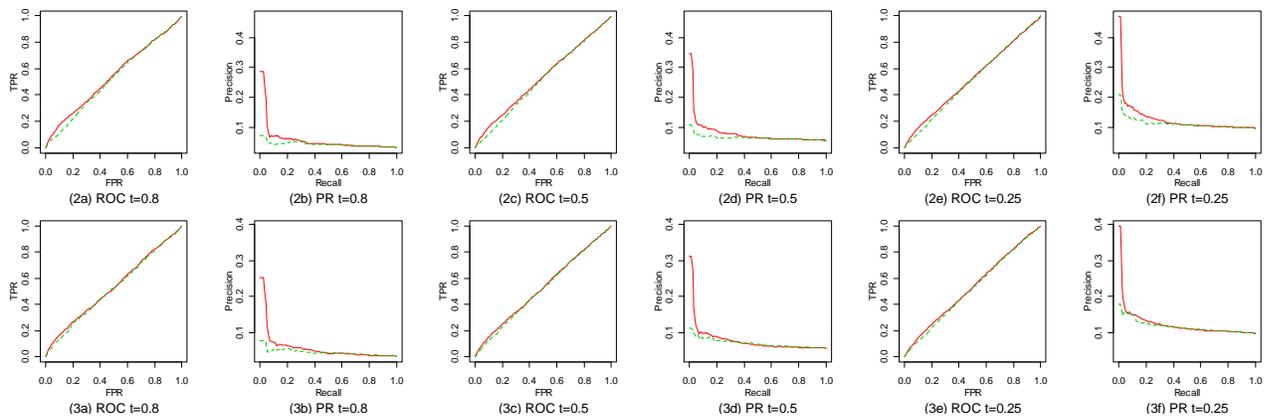

| | | | $t = 0.8$ | | | | $t = 0.5$ | | | | $t = 0.25$ | | | |
|---|---|---|---|---|---|---|---|---|---|---|---|---|---|---|
| | | | LIS | BH | AP | lfdr | LIS | BH | AP | lfdr | LIS | BH | AP | lfdr |
| $rr = 1.2$ | $\alpha = 0.05$ | FDR: | 0.018 | 0.059 | 0.059 | 0.010 | 0.018 | 0.059 | 0.059 | 0.010 | 0.018 | 0.058 | 0.058 | 0.009 |
| | | #TP: | 12 | 11 | 11 | 1 | 12 | 11 | 11 | 1 | 20 | 18 | 19 | 7 |
| | $\alpha = 0.10$ | FDR: | 0.077 | 0.089 | 0.089 | 0.010 | 0.077 | 0.089 | 0.089 | 0.010 | 0.076 | 0.079 | 0.079 | 0.009 |
| | | #TP: | 13 | 11 | 11 | 1 | 13 | 11 | 11 | 1 | 21 | 20 | 20 | 8 |
| $rr = 1.3$ | $\alpha = 0.05$ | FDR: | 0.047 | 0.044 | 0.054 | 0.015 | 0.047 | 0.044 | 0.064 | 0.005 | 0.046 | 0.044 | 0.064 | 0.014 |
| | | #TP: | 16 | 4 | 4 | 1 | 16 | 4 | 4 | 1 | 22 | 10 | 10 | 6 |
| | $\alpha = 0.10$ | FDR: | 0.067 | 0.104 | 0.104 | 0.015 | 0.067 | 0.104 | 0.104 | 0.005 | 0.066 | 0.103 | 0.103 | 0.014 |
| | | #TP: | 18 | 15 | 15 | 1 | 18 | 15 | 15 | 1 | 27 | 21 | 21 | 6 |

(1)

Figure 5: Comparison of BH, AP, localFDR and LIS in the additive models when we vary relative risk $rr$, $t$ and the nominal FDR level $\alpha$. Table (1) summarizes results. Subfigures (2a)-(2f) shows ROC and PR curves of LIS (solid red lines) and individual $p$-values (dashed green lines) with $rr = 1.2$. Subfigures (3a)-(3f) shows ROC and PR curves of LIS (solid red lines) and individual $p$-values (dashed green lines) with $rr = 1.3$.

over the baselines in the dominant model. The LIS procedure also recaptures almost the same number of true positives as BH and AP while maintaining a much lower FDR in the recessive model. Again, we further plot the ROC curves and precision-recall curves when we rank SNPs by LIS and by the $p$-values from individual tests. Subfigures (2a)-(2f) are for the dominant model. Subfigures (3a)-(3f) are for the recessive model. It is also observed that the curves from LIS dominate those from the $p$-values from individual tests in most places, which also suggests that LIS provides a more efficient ranking.

## 5 Real-world Application

Our primary GWAS dataset on breast cancer is from NCI's Cancer Genetics Markers of Susceptibility (CGEMS) (Hunter et al., 2007). $528,173$ SNPs for $1,145$ cases and $1,142$ controls are genotyped on the Illumina HumanHap500 array. Our secondary GWAS dataset comes from Marshfield Clinic. The Personalized Medicine Research Project (McCarty et al., 2005), sponsored by Marshfield Clinic, was used as the sampling frame to identify 162 breast cancer cases and 162 controls. The project was reviewed and approved by the Marshfield Clinic IRB. Subjects were selected using clinical data from the Marshfield Clinic Cancer Registry and Data Warehouse. Cases were defined as women having a confirmed diagnosis of breast cancer. Both the cases and controls had to have at least one mammogram within 12 months prior to having a biopsy. The subjects also had DNA samples that were genotyped using the Illumina HumanHap660 array, as part of the eMERGE (electronic MEdical Records and Genomics) network by McCarty et al. (2011).

We apply our multiple testing procedure on the CGEMS data. The settings of the procedure are the same as in the simulations on genetic data in Section 4. The individual test is two-proportion $z$-test. Our procedure reports 32 SNPs with LIS value of 0.0 (an estimated probability 1.0 of being associated). We further calculate the per-allele odds-ratio of these SNPs on the Marshfield data, and 14 of them have an odds-ratio around 1.2 or above. The details about the 14 SNPs are given in supplementary material. There are two clusters among them. First, rs3870371, rs7830137 and rs920455 (on chromosome 8) locate near each other and near the gene hyaluronan synthase 2 (HAS2) which has been shown to be associated with invasive breast cancer by many studies (Udabage et al., 2005;

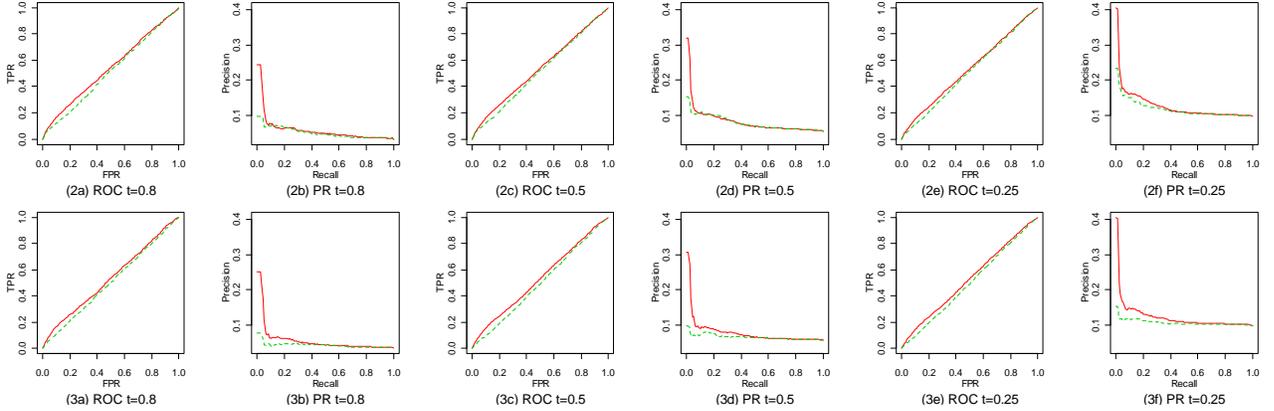

|  |  |  | $t = 0.8$ | | | | $t = 0.5$ | | | | $t = 0.25$ | | | |
|---|---|---|---|---|---|---|---|---|---|---|---|---|---|---|
|  |  |  | LIS | BH | AP | lfdr | LIS | BH | AP | lfdr | LIS | BH | AP | lfdr |
| **Dominant** | $\alpha = 0.05$ | FDR: | 0.026 | 0.040 | 0.040 | 0.010 | 0.026 | 0.040 | 0.040 | 0.010 | 0.025 | 0.039 | 0.039 | 0.009 |
|  |  | #TP: | 14 | 4 | 4 | 2 | 14 | 4 | 4 | 2 | 21 | 10 | 10 | 7 |
|  | $\alpha = 0.10$ | FDR: | 0.051 | 0.079 | 0.089 | 0.010 | 0.048 | 0.079 | 0.109 | 0.010 | 0.044 | 0.079 | 0.109 | 0.009 |
|  |  | #TP: | 20 | 12 | 12 | 3 | 22 | 12 | 12 | 3 | 33 | 19 | 29 | 18 |
| **Recessive** | $\alpha = 0.05$ | FDR: | 0.009 | 0.079 | 0.079 | 0.009 | 0.009 | 0.079 | 0.079 | 0.009 | 0.009 | 0.079 | 0.079 | 0.009 |
|  |  | #TP: | 11 | 11 | 11 | 11 | 11 | 11 | 11 | 11 | 18 | 17 | 18 | 17 |
|  | $\alpha = 0.10$ | FDR: | 0.018 | 0.104 | 0.104 | 0.009 | 0.018 | 0.104 | 0.114 | 0.009 | 0.017 | 0.104 | 0.114 | 0.009 |
|  |  | #TP: | 11 | 12 | 12 | 11 | 11 | 12 | 12 | 11 | 22 | 21 | 21 | 17 |

(1)

Figure 6: Comparison of BH, AP, localFDR and LIS in the dominant model and the recessive model with different $t$ values and different nominal FDR $\alpha$ values. Table (1) summarizes results. Subfigures (2a)-(2f) shows ROC and PR curves of LIS (solid red lines) and individual $p$-values (dashed green lines) in the dominant model. Subfigures (3a)-(3f) shows ROC and PR curves of LIS and individual $p$-values in the recessive model.

Li et al., 2007; Bernert et al., 2011). The other cluster includes rs11200014, rs2981579, rs1219648, and rs2420946 on chromosome 10. They are exactly the 4 SNPs reported by Hunter et al. (2007). Their associated gene FGFR2 is also well known to be associated with breast cancer. SNP rs4866929 on chromosome 5 is also very likely to be associated because it is highly correlated ($r^2$=0.957) with SNP rs981782 (not included in our data) which was identified from a much larger dataset (4,398 cases and 4,316 controls and a follow-up confirmation stage on 21,860 cases and 22,578 controls) by Easton et al. (2007).

## 6 Conclusion

In this paper, we use an MRF-coupled mixture model to leverage the dependence in multiple testing problems, and show the improved numerical performance on a variety of simulations and its applicability in a real-world GWAS problem. A theoretical question of interest is whether this graphical model based procedure is optimal in the sense that it has the smallest FNR among all the valid procedures. The optimality of the oracle procedure can be proved under the compound decision framework (Sun & Cai, 2007, 2009), as long as an exact inference algorithm exists or an approximate inference algorithm can be guaranteed to converge to the correct marginal probabilities. The asymptotic optimality of the data-driven procedure (the FNR yielded by the data-driven procedure approaches the FNR yielded by the oracle procedure as the number of tests $m \to \infty$) requires consistent estimates of the unknown parameters in the graphical models. Parameter learning in undirected models is more complicated than in directed models due to the normalization constant. To the best of our knowledge, asymptotic properties of parameter learning for hidden MRFs and MRF-coupled mixture models have not been investigated. Therefore, we cannot prove the asymptotic optimality of the data-driven procedure so far, although we can observe its close-to-oracle performance in the basic simulations.

### Acknowledgements

The authors acknowledge the support of the Wisconsin Genomics Initiative, NCI grant R01CA127379-01 and its ARRA supplement 3R01CA127379-03S1, NIGMS grant R01GM097618-01, NLM grant R01LM011028-01, NIEHS grant 5R01ES017400-03, eMERGE grant 1U01HG004608-01, NSF grant DMS-1106586 and the UW Carbone Cancer Center.

# References

Armitage, P. (1955). Tests for linear trends in proportions and frequencies. *BIOMETRICS*, **11**, 375C386.

Asuncion, A. U., Liu, Q., Ihler, A. T., & Smyth, P. (2010). Particle filtered MCMC-MLE with connections to contrastive divergence. In *ICML*.

Baum, L. E., Petrie, T., Soules, G., & Weiss, N. (1970). A maximization technique occurring in the statistical analysis of probabilistic functions of Markov chains. *ANN MATH STAT*, **41**(1), 164–171.

Benjamini, Y. & Heller, R. (2007). False discovery rates for spatial signals. *J AM STAT ASSOC*, **102**, 1272–1281.

Benjamini, Y. & Hochberg, Y. (1995). Controlling the false discovery rate: A practical and powerful approach to multiple testing. *J ROY STAT SOC B*, **57**(1), 289–300.

Benjamini, Y. & Hochberg, Y. (2000). On the adaptive control of the false discovery rate in multiple testing with independent statistics. *J EDUC BEHAV STAT*, **25**(1), 60–83.

Benjamini, Y. & Yekutieli, D. (2001). The control of the false discovery rate in multiple testing under dependency. *ANN STAT*, **29**, 1165–1188.

Bernert, B., Porsch, H., & Heldin, P. (2011). Hyaluronan synthase 2 (HAS2) promotes breast cancer cell invasion by suppression of tissue metalloproteinase inhibitor 1 (TIMP-1). *J BIOL CHEM*, **286**(49), 42349–42359.

Blanchard, G. & Roquain, E. (2009). Adaptive false discovery rate control under independence and dependence. *J MACH LEARN RES*, **10**, 2837–2871.

Celeux, G., Forbes, F., & Peyrard, N. (2003). EM procedures using mean field-like approximations for Markov model-based image segmentation. *Pattern Recognition*, **36**, 131–144.

Chatzis, S. P. & Varvarigou, T. A. (2008). A fuzzy clustering approach toward hidden Markov random field models for enhanced spatially constrained image segmentation. *IEEE Transactions on Fuzzy Systems*, **16**, 1351 – 1361.

Cochran, W. G. (1954). Some methods for strengthening the common chi-square tests. *BIOMETRICS*, **10**, 417–451.

Crouse, M. S., Nowak, R. D., & Baraniuk, R. G. (1998). Wavelet-based statistical signal processing using hidden Markov models. *IEEE T SIGNAL PROCES*, **46**(4), 886–902.

Easton, D. F., Pooley, K. A., Dunning, A. M., Pharoah, P. D. P., Thompson, D., Ballinger, D. G., Struewing, J. P., Morrison, J., Field, H., Luben, R., Wareham, N., Ahmed, S., Healey, C. S., Bowman, R., Meyer, K. B., Haiman, C. A., Kolonel, L. K., Henderson, B. E., Le Marchand, L., Brennan, P., Sangrajrang, S., Gaborieau, V., Odefrey, F., Shen, C.-Y., Wu, P.-E., Wang, H.-C., Eccles, D., Evans, G. D., Peto, J., Fletcher, O., Johnson, N., Seal, S., Stratton, M. R., Rahman, N., Chenevix-Trench, G., Bojesen, S. E., Nordestgaard, B. G., Axelsson, C. K., Garcia-Closas, M., Brinton, L., Chanock, S., Lissowska, J., Peplonska, B., Nevanlinna, H., Fagerholm, R., Eerola, H., Kang, D., Yoo, K.-Y., Noh, D.-Y., Ahn, S.-H., Hunter, D. J., Hankinson, S. E., Cox, D. G., Hall, P., Wedren, S., Liu, J., Low, Y.-L., Bogdanova, N., Schürmann, P., Dörk, T., Tollenaar, R. A. E. M., Jacobi, C. E., Devilee, P., Klijn, J. G. M., Sigurdson, A. J., Doody, M. M., Alexander, B. H., Zhang, J., Cox, A., Brock, I. W., Macpherson, G., Reed, M. W. R., Couch, F. J., Goode, E. L., Olson, J. E., Meijers-Heijboer, H., van den Ouweland, A., Uitterlinden, A., Rivadeneira, F., Milne, R. L., Ribas, G., Gonzalez-Neira, A., Benitez, J., Hopper, J. L., Mccredie, M., Southey, M., Giles, G. G., Schroen, C., Justenhoven, C., Brauch, H., Hamann, U., Ko, Y.-D., Spurdle, A. B., Beesley, J., Chen, X., Mannermaa, A., Kosma, V.-M., Kataja, V., Hartikainen, J., Day, N. E., Cox, D. R., & Ponder, B. A. J. (2007). Genome-wide association study identifies novel breast cancer susceptibility loci. *Nature*, **447**, 1087–1093.

Efron, B. (2007). Correlation and large-scale simultaneous significance testing. *J AM STAT ASSOC*, **102**(477), 93–103.

Efron, B., Tibshirani, R., Storey, J. D., & Tusher, V. (2001). Empirical Bayes analysis of a microarray experiment. *J AM STAT ASSOC*, **96**, 1151–1160.

Fan, J., Han, X., & Gu, W. (2012). Control of the false discovery rate under arbitrary covariance dependence. (to appear) J AM STAT ASSOC.

Farcomeni, A. (2007). Some results on the control of the false discovery rate under dependence. *SCAND J STAT*, **34**(2), 275–297.

Finner, H. & Roters, M. (2002). Multiple hypotheses testing and expected number of type I errors. *ANN STAT*, **30**, 220–238.

Freidlin, B., Zheng, G., Li, Z., & Gastwirth, J. L. (2002). Trend tests for case-control studies of genetic markers: power, sample size and robustness. *HUM HERED*, **53**(3), 146–152.

Friguet, C., Kloareg, M., & Causeur, D. (2009). A factor model approach to multiple testing under dependence. *J AM STAT ASSOC*, **104**(488), 1406–1415.

Ganapathi, V., Vickrey, D., Duchi, J., & Koller, D.

(2008). Constrained approximate maximum entropy learning of Markov random fields. In *UAI*.

Gelfand, A. E. & Smith, A. F. M. (1990). Sampling-based approaches to calculating marginal densities. *J AM STAT ASSOC*, **85**(410), 398–409.

Genovese, C. & Wasserman, L. (2002). Operating characteristics and extensions of the false discovery rate procedure. *J ROY STAT SOC B*, **64**, 499–517.

Genovese, C. & Wasserman, L. (2004). A stochastic process approach to false discovery control. *ANN STAT*, **32**, 1035–1061.

Genovese, C., Roeder, K., & Wasserman, L. (2006). False discovery control with p-value weighting. *BIOMETRIKA*, **93**, 509–524.

Geyer, C. J. (1991). Markov chain Monte Carlo maximum likelihood. *COMP SCI STAT*, pages 156–163.

Hinton, G. (2002). Training products of experts by minimizing contrastive divergence. *NEURAL COMPUT*, **14**, 1771–1800.

Hunter, D. J., Kraft, P., Jacobs, K. B., Cox, D. G., Yeager, M., Hankinson, S. E., Wacholder, S., Wang, Z., Welch, R., Hutchinson, A., Wang, J., Yu, K., Chatterjee, N., Orr, N., Willett, W. C., Colditz, G. A., Ziegler, R. G., Berg, C. D., Buys, S. S., Mccarty, C. A., Feigelson, H. S., Calle, E. E., Thun, M. J., Hayes, R. B., Tucker, M., Gerhard, D. S., Fraumeni, J. F., Hoover, R. N., Thomas, G., & Chanock, S. J. (2007). A genome-wide association study identifies alleles in FGFR2 associated with risk of sporadic postmenopausal breast cancer. *NAT GENET*, **39**(7), 870–874.

Jordan, M. I., Ghahramani, Z., Jaakkola, T., & Saul, L. K. (1999). An introduction to variational methods for graphical models. *MACH LEARN*, **37**, 183–233.

Kschischang, F., Frey, B., & Loeliger, H.-A. (2001). Factor graphs and the sum-product algorithm. *IEEE T INFORM THEORY*, **47**(2), 498 –519.

Lauritzen, S. L. & Spiegelhalter, D. J. (1988). Local computations with probabilities on graphical structures and their application to expert systems. *J ROY STAT SOC B*, **50**(2), 157–224.

Leek, J. T. & Storey, J. D. (2008). A general framework for multiple testing dependence. *P NATL ACAD SCI USA*, **105**(48), 18718–18723.

Li, Y., Li, L., Brown, T. J., & Heldin, P. (2007). Silencing of hyaluronan synthase 2 suppresses the malignant phenotype of invasive breast cancer cells. *INT J CANCER*, **120**(12), 2557–2567.

McCarty, C., Wilke, R., Giampietro, P., Wesbrook, S., & Caldwell, M. (2005). Marshfield Clinic Personalized Medicine Research Project (PMRP): design, methods and recruitment for a large population-based biobank. *PERS MED*, **2**, 49–79.

McCarty, C. A., Chisholm, R. L., Chute, C. G., Kullo, I. J., Jarvik, G. P., Larson, E. B., Li, R., Masys, D. R., Ritchie, M. D., Roden, D. M., Struewing, J. P., Wolf, W. A., & eMERGE Team (2011). The eMERGE Network: a consortium of biorepositories linked to electronic medical records data for conducting genomic studies. *BMC MED GENET*, **4**(1), 13.

Murphy, K. P., Weiss, Y., & Jordan, M. I. (1999). Loopy belief propagation for approximate inference: An empirical study. In *UAI*, pages 467–475.

Owen, A. B. (2005). Variance of the number of false discoveries. *J ROY STAT SOC B*, **67**, 411–426.

Romano, J., Shaikh, A., & Wolf, M. (2008). Control of the false discovery rate under dependence using the bootstrap and subsampling. *TEST*, **17**, 417–442.

Salakhutdinov, R. (2009). Learning in Markov random fields using tempered transitions. In *NIPS*, pages 1598–1606.

Sarkar, S. K. (2006). False discovery and false nondiscovery rates in single-step multiple testing procedures. *ANN STAT*, **34**(1), 394–415.

Schraudolph, N. N. (2010). Polynomial-time exact inference in NP-hard binary MRFs via reweighted perfect matching. In *AISTATS*.

Schraudolph, N. N. & Kamenetsky, D. (2009). Efficient exact inference in planar Ising models. In *NIPS*.

Slager, S. L. & Schaid, D. J. (2001). Case-control studies of genetic markers: power and sample size approximations for Armitage's test for trend. *HUM HERED*, **52**(3), 149–153.

Storey, J. D. (2002). A direct approach to false discovery rates. *J ROY STAT SOC B*, **64**, 479–498.

Storey, J. D. (2003). The positive false discovery rate: A Bayesian interpretation and the q-value. *ANN STAT*, **31**(6), 2013–2035.

Su, Z., Marchini, J., & Donnelly, P. (2011). HAPGEN2: simulation of multiple disease SNPs. *BIOINFORMATICS*.

Sun, W. & Cai, T. T. (2007). Oracle and adaptive compound decision rules for false discovery rate control. *J AM STAT ASSOC*, **102**(479), 901–912.

Sun, W. & Cai, T. T. (2009). Large-scale multiple testing under dependence. *J ROY STAT SOC B*, **71**, 393–424.

The International HapMap Consortium (2003). The international HapMap project. *NATURE*, **426**, 789–796.

Tieleman, T. (2008). Training restricted Boltzmann machines using approximations to the likelihood gradient. In *ICML*, pages 1064–1071.

Tieleman, T. & Hinton, G. (2009). Using fast weights to improve persistent contrastive divergence. In *ICML*, pages 1033–1040.

Udabage, L., Brownlee, G. R., Nilsson, S. K., & Brown, T. J. (2005). The over-expression of HAS2, Hyal-2 and CD44 is implicated in the invasiveness of breast cancer. *EXP CELL RES*, **310**(1), 205 – 217.

Wainwright, M. J., Jaakkola, T. S., & Willsky, A. S. (2003a). Tree-based reparameterization framework for analysis of sum-product and related algorithms. *IEEE T INFORM THEORY*, **49**, 2003.

Wainwright, M. J., Jaakkola, T. S., & Willsky, A. S. (2003b). Tree-reweighted belief propagation algorithms and approximate ML estimation via pseudo-moment matching. In *AISTATS*.

Weiss, Y. (2000). Correctness of local probability propagation in graphical models with loops. *NEURAL COMPUT*, **12**(1), 1–41.

Welling, M. & Sutton, C. (2005). Learning in Markov random fields with contrastive free energies. In *AISTATS*.

Wu, M. C., Kraft, P., Epstein, M. P., Taylor, D. M., Chanock, S. J., Hunter, D. J., & Lin, X. (2010). Powerful SNP-set analysis for case-control genome-wide association studies. *AM J HUM GENET*, **86**(6), 929–942.

Wu, W. B. (2008). On false discovery control under dependence. *ANN STAT*, **36**(1), 364–380.

Yedidia, J. S., Freeman, W. T., & Weiss, Y. (2000). Generalized belief propagation. In *NIPS*, pages 689–695. MIT Press.

Yekutieli, D. & Benjamini, Y. (1999). Resampling-based false discovery rate controlling multiple test procedures for correlated test statistics. *J STAT PLAN INFER*, **82**, 171–196.

Zhang, C., Fan, J., & Yu, T. (2011). Multiple testing via $FDR_L$ for large-scale imaging data. *ANN STAT*, **39**(1), 613–642.

Zhang, Y., Brady, M., & Smith, S. (2001). Segmentation of brain MR images through a hidden Markov random field model and the expectation-maximization algorithm. *IEEE Transactions on Medical Imaging*.